# On-demand steering of hyperbolic chiral polaritons

Andrea S. Dai[1,#], Fuyang Tay[1,2,#], Ding Xu[1], Inki Lee[1], Noah Bussell[1], Daria Balatsky[3,4], Francesco L. Ruta[2,5], Emma Lian[1], Colin Nuckolls[1], Xavier Roy[1], James G. Analytis[4,6,7], Andrew J. Millis[2,8], D. N. Basov[2*], Milan Delor[1*]

[1]Department of Chemistry, Columbia University, New York, NY, USA.
[2]Department of Physics, Columbia University, New York, NY, USA.
[3]Department of Chemistry, University of California, Berkeley, CA, USA.
[4]Department of Physics, University of California, Berkeley, CA, USA.
[5]Department of Applied Physics and Applied Mathematics, Columbia University, New York, NY, USA.
[6]CIFAR Quantum Materials, Toronto, ON, Canada.
[7]Kavli Energy Nanoscience Institute, Berkeley, CA, USA.
[8]Center for Computational Quantum Physics, Flatiron Institute, New York, NY, USA.

[#]These authors contributed equally to this work.

[*]Corresponding authors. db3056@columbia.edu; milan.delor@columbia.edu

## Abstract

Control of light polarization and propagation in sub-wavelength architectures is foundational to nanophotonic technologies. A frontier direction is to leverage strong optical spin-orbit interactions to realize polarization-selective light steering, known as the photonic spin Hall effect. In this context, hyperbolic plasmon polaritons (HPPs) are of particular interest as they offer large optical spin-orbit coupling from strong confinement and dielectric anisotropy, as well as ray-like propagation. Despite theoretical predictions, however, the hyperbolic spin Hall effect in natural materials has remained elusive. Here, we demonstrate the hyperbolic spin Hall effect in the visible and near-infrared range in the natural hyperbolic van der Waals metal $MoOCl_2$. Enabling this discovery is a novel far-field pump-probe microscope that facilitates the launching and imaging of HPPs with exceptional sensitivity through interference with a high-momentum reference field. This approach preserves excellent control over light polarization, overcoming a key barrier to polarization-selective interrogation of hyperbolic materials. We show that both hyperbolic and surface plasmons in $MoOCl_2$ display chiral fields, and that their propagation direction can be completely switched upon light helicity reversal. Our results demonstrate on-demand steering of chiral plasmons, firmly establishing natural hyperbolic materials as ideal components for reconfigurable nanophotonics and chiral light-matter coupling.

### Introduction

Emerging photonic and optoelectronic technologies require exquisite control over light propagation and polarization while confining light to ever-smaller scales. Hyperbolic polaritons (HPs) in van der Waals (vdW) materials have received significant interest in this context as they combine deep subwavelength confinement with highly directional, ray-like propagation[1–3]. HPs emerge in highly anisotropic materials with dielectric functions of opposing signs along different crystal axes. Such confinement and anisotropy are also key ingredients for large optical spin–orbit interactions[4,5] and associated helicity–momentum locking, a sought-after effect where the polariton handedness is directly linked to its trajectory. While this photonic spin Hall effect (SHE) is documented in structured photonic media and dielectric cavities[6–14], it remains elusive in natural hyperbolic materials – systems that offer the ultimate platform for optical spin-orbit interactions due to theoretically unbounded light confinement. Realizing the hyperbolic SHE would enable on-demand control over deeply subdiffractional ray propagation through helicity-dependent excitation[15–17] (Fig. 1a), providing a path toward subwavelength logical gates and the transport of spin-encoded information in nanophotonics.

Beyond identifying suitable hyperbolic materials with strong optical spin-orbit interactions, a central obstacle to the discovery of the hyperbolic SHE is the large momentum mismatch between free-space light and HPs. Indeed, HPs have been studied primarily through specialized near-field optics[18–23] and electron microscopy[24–27] due to their unique capacity to interrogate deeply subdiffractional (high-momentum) modes; however, these methods possess limited control over polarization. Developing a general approach to access and control HPs using far-field optics with excellent polarization control is thus strongly desired. The challenge is that a far-field objective's passband is limited to $k_{\mathrm{max}} = k_0\mathrm{NA}$, where $k_0 = \frac{2\pi}{\lambda}$ is the free-space wavevector bounded by the vacuum light cone, and NA is the numerical aperture of the objective. This passband is insufficient for accessing confined high-$k$ hyperbolic modes. Nevertheless, for coherent imaging modalities, Abbe demonstrated in 1873[28] that this passband can be doubled using oblique illumination to shift high-order diffraction modes into the aperture—an aperture-engineering principle similar to structured illumination used in incoherent (fluorescence) superresolution microscopy[29–32].

Here, we develop oblique-illumination pump-probe microscopy to access in-plane hyperbolic plasmon polaritons (HPPs) in coherent far-field imaging. Leveraging our full control over light momentum, energy, and polarization, we image and control the propagation of visible- and near-infrared surface plasmons and HPPs in $MoOCl_2$. $MoOCl_2$ is a naturally occurring vdW material with strong optical anisotropy arising from an orbital-selective Peierls distortion: the $a$-axis is metallic, with bands crossing the Fermi level, while the $b$-axis bands split away from it[21,22,25,33–37] (Fig. 1b). This anisotropy is conveyed in a permittivity tensor with opposite signs along different crystal axes over a broad frequency range ($\varepsilon_a < 0$; $\varepsilon_b, \varepsilon_c > 0$; Fig. 1c). Through imaging and simulations, we show that HPPs in $MoOCl_2$ are intrinsically chiral and display strong helicity-momentum locking. We demonstrate excitation of unidirectional HPPs with circularly polarized light, with propagation direction entirely controlled by light helicity[15–17]. We replicate this result for both hyperbolic and elliptical plasmons that co-exist in $MoOCl_2$. Realizing the sought-after hyperbolic SHE crucially advances our ability to control beam steering, polarization, and nonlinearities in vdW platforms.

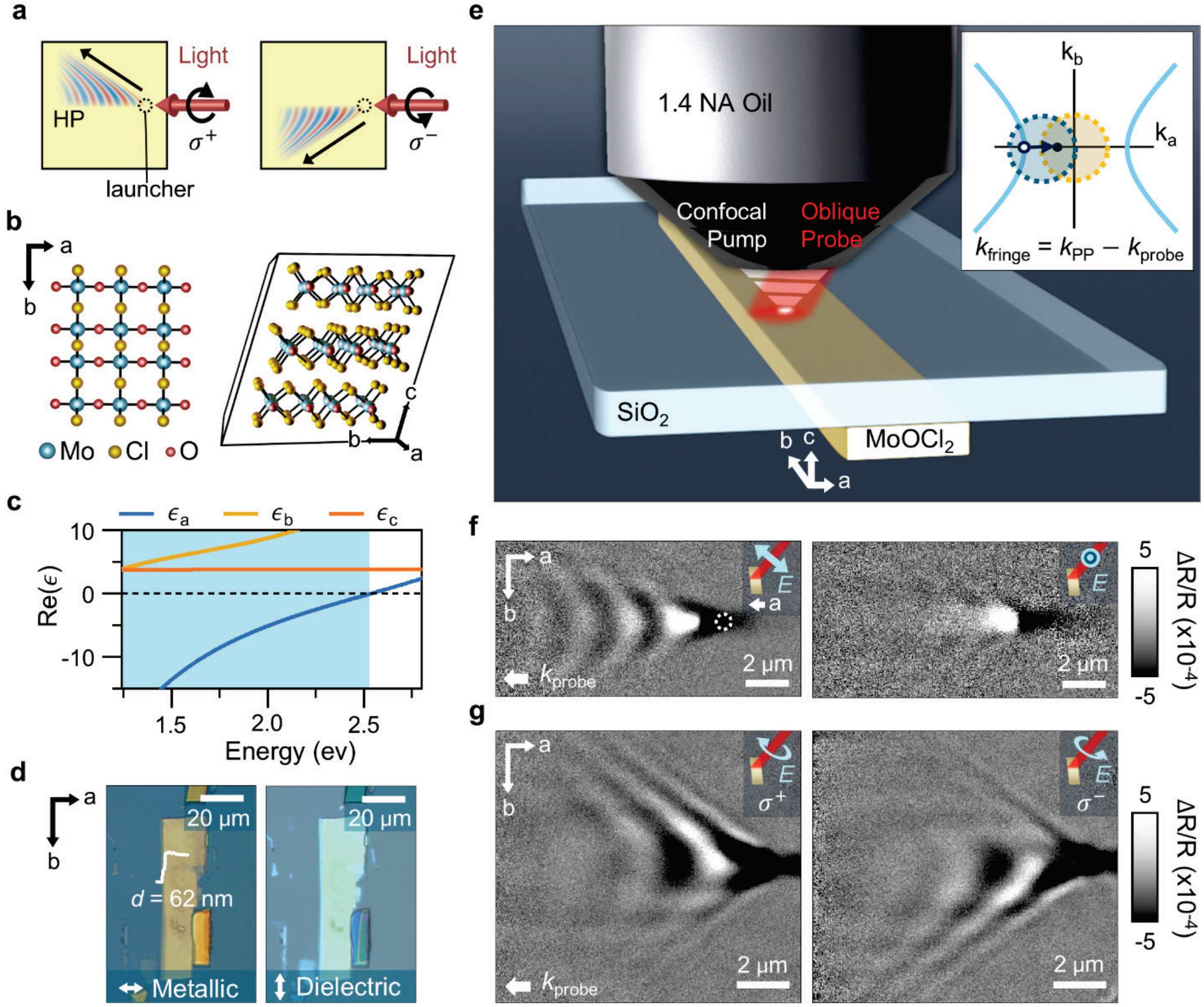

**Fig. 1. Probing hyperbolic plasmon polaritons in $MoOCl_2$ via optical pump-probe microscopy. a,** Conceptual schematic of helicity-dependent launching of hyperbolic polaritons (HPs): opposite light helicities ($\sigma^+$, $\sigma^-$) selectively excite upper and lower rays. **b,** Crystal structure of $MoOCl_2$. **c,** Real part of the dielectric function, Re($\varepsilon$), of $MoOCl_2$. The shaded region represents the hyperbolic range. **d,** Optical images of an exfoliated 62-nm-thick $MoOCl_2$ flake on $SiO_2$ under linearly polarized illumination along the *a* (metallic) and *b* (dielectric) axes. **e,** Schematic diagram of pump-probe microscopy with confocal pump and oblique probe, tilted along the *a*-axis of the $MoOCl_2$ crystal. Inset: schematic depicting oblique illumination shifting the detectable passband (yellow circle) toward higher in-plane momentum along the metallic axis, $k_a$ (blue circle), overlapping with a high *k* HP mode. **f,** Pump-probe images of the crystal in **d** at 0.4 ns delay, $k_{probe}$ ~NA $k_0$, and pump/probe energies of 1.72 eV/1.63 eV with oblique incidence probe for p-polarization ($E /\!/ a,c$, left) and s-polarization ($E /\!/ b$, right). The white circle indicates the pump excitation position. **g,** Pump-probe images acquired with circularly polarized probe beams ($\sigma^+$ & $\sigma^-$) at 0.4 ns delay, $k_{probe}$ ~NA $k_0$, and pump/probe energies of 1.72 eV/1.80 eV.

**Helicity-controlled unidirectional launching of hyperbolic polaritons**

We first perform measurements on a mechanically exfoliated 62-nm-thick $MoOCl_2$ microcrystal residing on a $SiO_2$ substrate. Under steady-state optical microscopy, the crystal shows pronounced polarization-dependent contrast (Fig. 1d), offering visual evidence of in-plane dielectric anisotropy: one polarization yields metallic reflectance, while the orthogonal polarization displays dielectric (transparent) behavior. To launch and image plasmons, we employ pump-probe interferometric microscopy with a diffraction-limited, focused pump beam, followed by a widefield probe beam obliquely incident on the crystal from the substrate side[38] (Fig. 1e). The backscattered light, which carries the signal of interest, interferes with the reflected field at the sample-substrate interface, which acts as a reference field[39,40]. An oil-immersion objective with NA = 1.4 is used to provide access to high in-plane probe wavevectors, $k_{\text{probe}}$. The focused pump beam locally and transiently alters the plasma frequency and damping in $MoOCl_2$ through carrier photoexcitation and heating, inducing a local refractive-index mismatch lasting for a few nanoseconds. The mismatched region acts as a pump-induced scatterer that couples the time-delayed probe into plasmons in $MoOCl_2$; this approach is equivalent to placing an artificial launcher like a metal disk or nanoparticle on the material[21,41,42], but is fully reversible and requires no fabrication. Importantly, pump modulation turns this scattering region 'on' and 'off' at 1 kHz; differential pump on/pump off images provide shot-noise-limited images of plasmon spatial profiles with an exceptional sensitivity to refractive index changes of $10^{-6}$.

The obliquely incident probe field is the key to accessing HPPs with wavevectors $k_{\text{PP}}$ beyond the objective NA, as it shifts the effective passband of the objective to higher wavevectors through interferometric downshifting (Fig. 1e inset, yellow circle for normal incidence and blue circle for oblique incidence). For example, the pump-probe image in Figure 1f (left) exhibits periodic fringe patterns arising from interference between the field scattered off of plasmons in $MoOCl_2$, and the reflected probe field at $k_{\text{probe}}$. The fringe wavevector is then given by $k_{\text{fringe}} = k_{\text{PP}} \pm k_{\text{probe}}$. The subtractive term downshifts high-$k$ plasmon fringes into the objective's passband, enabling detection of features with in-plane momenta up to 2NA $k_0$.

Figure 1f displays pump-probe images collected with a probe energy of 1.63 eV. $k_{\text{probe}}$ is close to the NA limit and is linearly polarized along either the crystal *a* (metallic) or *b* (dielectric) axes. For polarization $E \parallel a$, pronounced 'V'-shaped periodic features appear, closely resembling those observed with near-field and electron microscopy images of HPPs in $MoOCl_2$[21,22,26,27]. These hyperbolic fringes are observed only on the forward side of the probe (along $+k_{\text{probe}}$, left side of the scatterer), whereas the backward direction ($-k_{\text{probe}}$, right side) is dominated by the point spread function (PSF) under oblique incidence[43,44] (Supplementary Sections 1-4). The signal is suppressed for $E \parallel b$. Furthermore, the 'V'-shaped periodic features are only observable when the probe is tilted along the metallic *a*-axis (Supplementary Fig. 2), supporting the assignment of these fringes to HPPs. We return to a comprehensive momentum- and energy-dependent analysis of the HPP and surface plasmon excitations below.

Our key observation of the hyperbolic SHE is displayed in Figure 1g, where we use circularly polarized probe pulses to launch HPPs in $MoOCl_2$. Figure 1g compares images acquired with opposite light helicities, $\sigma^+$ (clockwise) and $\sigma^-$ (counterclockwise) polarization. We observe clear helicity–momentum locking: the handedness of the light launching the HPPs controls the propagation direction of the polariton rays, with $\sigma^+$ selectively exciting the upper polariton ray, and $\sigma^-$ selectively launching the lower polariton ray. These results clearly demonstrate the natural hyperbolic SHE, providing a new route to manipulate the propagation of deeply confined polaritons in anisotropic vdW crystals via polarization.

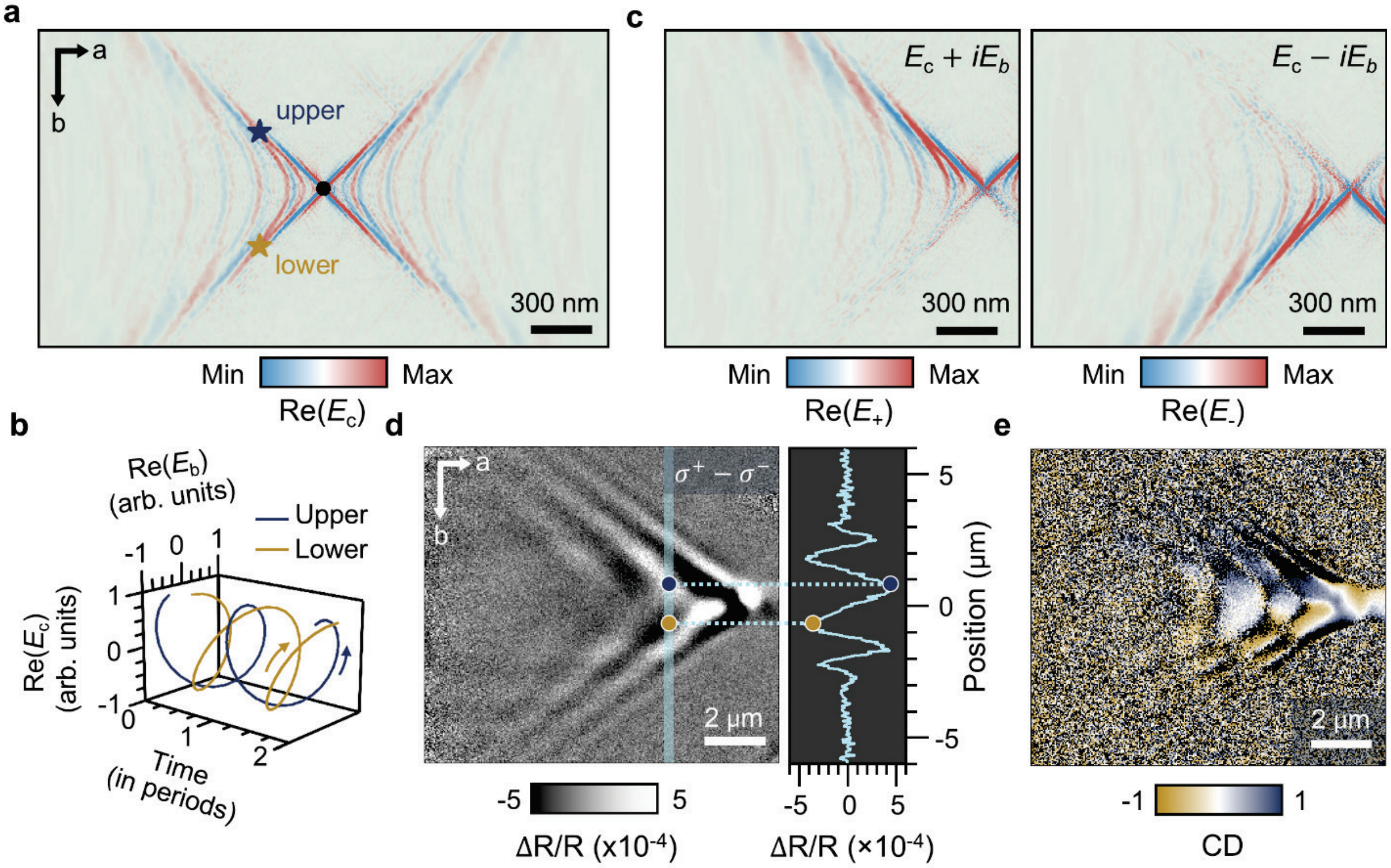

**Fig. 2. Chiral polaritons and helicity-dependent polariton launching in $MoOCl_2$. a,** Simulated out-of-plane electric field distribution, Re($E_c$), for a plane wave scattered by a cylindrical defect inside $MoOCl_2$ (see Methods). The black dot denotes the position of the defect region. **b,** Time evolution of the electric field components in the *bc* plane at the positions of the upper and lower plasmon-polariton rays indicated by the stars in **a**, revealing opposite helicities. **c,** Circularly polarized components of the simulated electric field in the *bc* plane with opposite helicities, $E_\pm = E_c \pm iE_b$. **d,** Difference between the pump-probe images acquired with opposite probe helicities (from Fig. 1g). The right panel shows the line profile extracted across the shaded region. **e,** Circular dichroism (CD) map of the same pump-probe images from Figure 1g, where CD = $(R_{\sigma^+} - R_{\sigma^-})$ / $(R_{\sigma^+} + R_{\sigma^-})$.

To elucidate the origin of this helical selectivity, Figure 2a displays the simulated real part of the out-of-plane electric field, $E_c$, for a linearly polarized plane wave scattered by a cylindrical inhomogeneity in $MoOCl_2$. The cylindrical region possesses a locally modified plasma frequency, emulating the pump-induced scatterer in the experiments (see Methods). The simulation shows the excitation of two hyperbolic rays (upper and lower). Notably, the time evolution of the electric fields at the marked positions (stars in Fig. 2a) shows that the fields for the upper and lower rays rotate in opposite directions (Fig. 2b). These simulations thus indicate opposite polariton helicities for the upper and lower hyperbolic rays. Furthermore, Figure 2c plots the calculated electric field distributions of the scattered field for $E_\pm = E_c \pm iE_b$, corresponding to *bc*-plane circular polarization. The simulations confirm that opposite hyperbolic rays possess opposite circular polarizations. The combination of helical field evolution and helicity-dependent propagation are hallmarks of chiral polaritons and the hyperbolic spin Hall effect.

A convenient representation of the helicity-dependent propagation is shown in Figure 2d. Here, we plot the difference of the two experimental pump-probe images in Figure 1g, i.e. $\sigma^+ - \sigma^-$, showing opposite interferometric signal for the two branches. The line profile in the inset shows nearly equal-magnitude but opposite-sign responses for upper and lower branches, indicating comparable launching efficiency for the two helicities. We also plot the effective circular dichroism, defined as CD = $(R_{\sigma^+} - R_{\sigma^-}) / (R_{\sigma^+} + R_{\sigma^-})$, in Figure 2e. Although noisy due to near-0 values in the denominator when the interferometric signal flips phase, we observe near-unity and opposite circular dichroism for the two hyperbolic rays, indicating near-complete selectivity of ray launching through polarization control.

**Tracking the transition from elliptical to hyperbolic plasmons**

$MoOCl_2$ exhibits an exceptionally rich plasmonic topology due to its biaxial hyperbolicity, which we now analyze in more detail. Figure 3a displays the calculated imaginary part of the reflection coefficient, Im($r_p$), at 1.63 eV for a 62 nm $MoOCl_2$ flake on a $SiO_2$ substrate. The calculated reflectance displays three plasmon-polariton (PP) regimes along the $a$-axis wavevector $k_a$: (i) vacuum surface plasmon polaritons (v-SPPs) between the vacuum and $SiO_2$ light lines; (ii) Surface plasmon polaritons (s-SPPs) between the $SiO_2$ and $MoOCl_2$ $c$-axis light lines; and (iii) HPPs extending to large $k_a$[21,22,25–27]. The SPPs are consistent with the directional leaky modes with lenticular dispersion reported in other anisotropic systems[21,45].

The momentum selectivity of oblique pump-probe microscopy allows systematically mapping these plasmonic modes across a broad $k$ range. As shown in Fig. 3b, the real-space fringe patterns transition from closed to open contours as $k_{probe}$ increases (see Supplementary Section 3 for quantification of $k_{probe}$). Under normal incidence ($k_{probe} \sim 0$), the pump-probe image displays an isotropic feature centered at the pump position. With increasing $k_{probe}$, the forward and backward portions of the signal become asymmetric ($k_{probe} = 0.96\ k_0$). A contrast phase reversal occurs as $k_{probe}$ crosses the vacuum light line (from $k_{probe} = 0.96\ k_0$ to $1.13\ k_0$). At larger $k_{probe}$, pronounced hyperbolic fringes emerge on the forward side ($k_{probe} = 1.39\ k_0$). This evolution of the plasmonic response matches the modal structure calculated in Figure 3a. Under normal incidence, the microscope passband primarily covers the v-SPP regime (yellow circle in Fig. 3a). For increasingly tilted probe fields, particularly near the NA limit, the corresponding shifted passband captures not only v-SPP but also s-SPP and HPP branches (blue circle in Fig. 3b). We therefore attribute the closed elliptical fringes at low $k_{probe}$ predominantly to v-SPPs, whereas the open-surface fringes observed at $k_{probe} \sim$ NA $k_0$ reflect contributions from multiple PP branches, including HPPs that are not detectable under normal incidence.

To quantitatively compare the experiments against the calculated dispersion, we Fourier transform the images for each incidence angle. Since $k_{fringe} = k_{PP} - k_{probe}$, we shift the images in Fourier space by adding $k_{probe}$ along $k_a$, recovering $k_{PP}$. Figure 3c displays the resulting frequency-shifted transforms (Supplementary Section 4), which trace the evolution of the isofrequency contours from closed (elliptical) to open (hyperbolic) surfaces. At $k_{probe} = 0.96\ k_0$, the Fourier-transformed data reveals an elliptical contour that matches the simulated v-SPP. Increasing $k_{probe}$ to $1.13\ k_0$ resolves the s-SPP next to the NA line. At $k_{probe} = 1.39\ k_0$, the data displays an open, hyperbolic contour consistent with the HPP surface. Note that across the working range of $k_{\text{probe}}$, the pump-probe signal and PSF have distinct positions in reciprocal space (Supplementary Fig. 4c) allowing deconvolution via Fourier filtering (Supplementary Fig. 5a-b). The results in Figure 3 confirm our ability to track the topological evolution from elliptical to hyperbolic plasmons in $MoOCl_2$.

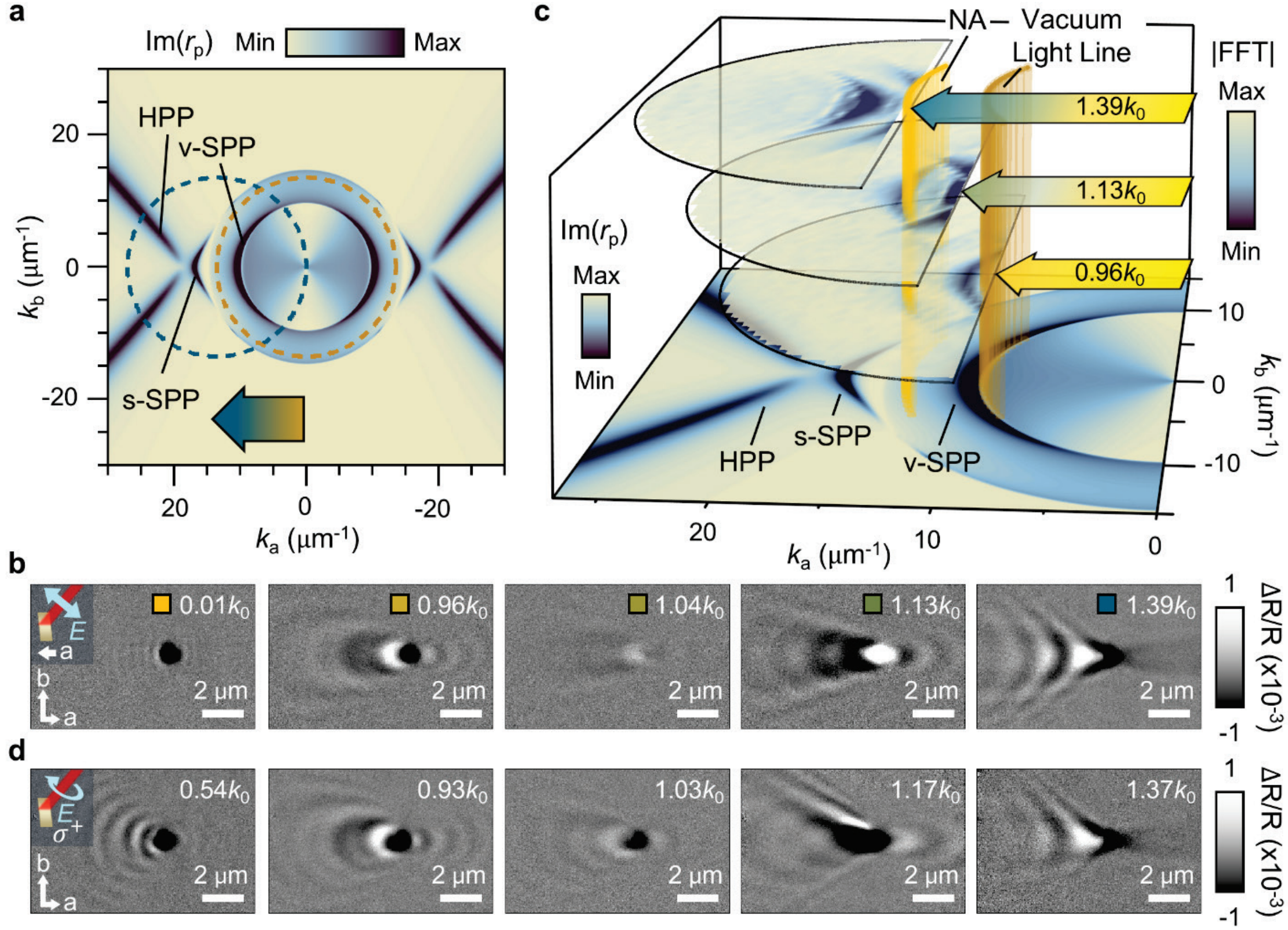

**Fig. 3. Mapping the evolution from elliptical to hyperbolic plasmons through momentum-dependent imaging. a,** Calculated dispersion at 1.91 eV for a $MoOCl_2$ flake ($d$ = 62 nm) on $SiO_2$. The vacuum (v-SPP) and $SiO_2$ (s-SPP) surface plasmon polaritons and hyperbolic plasmon polariton (HPP) are labeled. Dashed circles represent the probe passband under normal incidence (centered at $k_a$ = 0, yellow) and oblique incidence near the NA limit (offset to finite $k_a$, blue). **b,** Pump-probe images of the same flake at 0.4 ns delay for various $k_{\text{probe}}$. $k_0$ denotes the wavevector in free space. The pump/probe energies are 1.72 eV/ 1.91 eV. **c,** Processed Fourier transform (Supplementary Section 4) of **b.** The images are vertically offset for clarity. The calculated dispersion in **a** is projected at the bottom. The vacuum light line and the NA cutoff are vertically projected. **d,** Pump-probe images acquired with a $\sigma^+$-polarized probe for different $k_{\text{probe}}$.

Although HPPs display strong helicity-momentum locking, it is an open question whether the large anisotropy in $MoOCl_2$ provides sufficient optical spin-orbit interactions to support spin Hall effects for less confined plasmon modes. We address this question using our momentum selectivity. Figure 3d characterizes the degree of helical selectivity using a $\sigma^+$ probe for different incident wavevectors. The upper ray is launched with higher efficiency at all $k_{\text{probe}}$, indicating that both SPPs and HPPs exhibit some degree of helicity-momentum locking. However, the asymmetry between upper and lower rays is clearly largest for the hyperbolic rays (Figure 3d, right). This observation is consistent with the notion that optical spin-orbit interactions are enhanced under larger confinement[5], making HPPs ideal for the photonic SHE.

**Mapping plasmonic dispersions**

Figure 4a depicts a set of pump-probe images taken at various probe energies for the 62-nm-thick $MoOCl_2$ flake at $k_{probe} = 1.38k_0$. 2.48 eV is at the upper limit of the hyperbolic range (Fig. 1c) with high losses, and accordingly we do not observe hyperbolic fringes. For other probe energies, we fit the line profile (dashed line in Fig. 4a) with an exponentially damped sine function to extract $k_{fringe}$, which can equivalently be extracted through Fourier analysis (Supplementary Section 5). Using $k_{PP} = k_{fringe} + k_{probe}$, we plot $k_{PP}$ as a function of energy in Fig. 4b for both $k_{probe} = 0.76k_0$ (cyan circles) and $1.38k_0$ (magenta circles), overlaid with the calculated dispersion. Concurring with pump-probe images displaying elliptical versus hyperbolic scattering contours, the dispersions for $k_{probe} < k_0$ and $k_{probe} > k_0$ correspond to different plasmon branches. The former matches the calculated v-SPP dispersion across all measured energies. The values are also in excellent agreement with the plasmon dip measured in broadband angle-resolved reflectance using the same high-NA objective (left side of Fig. 4b).

For $k_{probe} = 1.38k_0$, the measured $k_{PP}$ overlaps well with the HPP mode particularly for lower probe energies (magenta circles in Fig. 4b), further supporting that the open hyperbolic contours stem from HPPs. Characteristic bending of the HPP dispersion is observed in the experiments, but is less pronounced than in the calculated dispersion, resulting in a lower $k_{HPP}$ than predicted at high energies. We offer two possible hypotheses for this deviation. First, dielectric functions are notoriously difficult to extract when approaching high-loss regions, such that the calculated HPP dispersion in Fig. 4b may not reflect the true HPP dispersion in $MoOCl_2$. Several dielectric functions for $MoOCl_2$ have been reported[21,22,34,37], and scattering-type scanning near-field optical microscopy (s-SNOM) measurements also exhibit deviations from simulated dispersions[21,22,37]. In addition, reported s-SNOM data is only available for energies up to 1.77 eV[21,22,37], precluding independent validation of HPP dispersions in this high-energy range where deviations occur. Among available dielectric functions, our far-field measurements agree best with the initially-reported dielectric function from density functional theory (DFT)[21,34]. A second possibility is that the deviation reflects interactions between HPPs and SPPs under our experimental conditions, resulting in leaky modes that deviate from the pure hyperbolic dispersion[45]. Future complementary measurements comparing s-SNOM and far-field measurements in this high-energy range will elucidate this deviation.

To calculate the quality ($Q$)-factor of the HPPs, we fit an exponentially-damped sinusoid to the fringes to extract the real and imaginary components of $k_{fringe}$. The $Q$-factor, defined as $\mathrm{Re}(k_{fringe})/\mathrm{Im}(k_{fringe})$, is plotted in Fig. 4c (solid circles). The measured $Q$-factors are in excellent agreement with the dispersion calculated from a recently-measured dielectric function[37] (open circles in Fig. 4c). In addition, thinner flakes show more confined HPPs and lower $Q$-factors (Supplementary Section 6), in agreement with the trend expected for HPP dispersions. However, unlike s-SNOM, highly confined HPPs in thin flakes are not accessible given our $k$-range limitation to $k_{probe} < 2\mathrm{NA}k_0$. We anticipate that future developments of coherent oblique-illumination microscopy using nonlinear enhancements[46] will enable accessing even more confined polaritons in the far field.

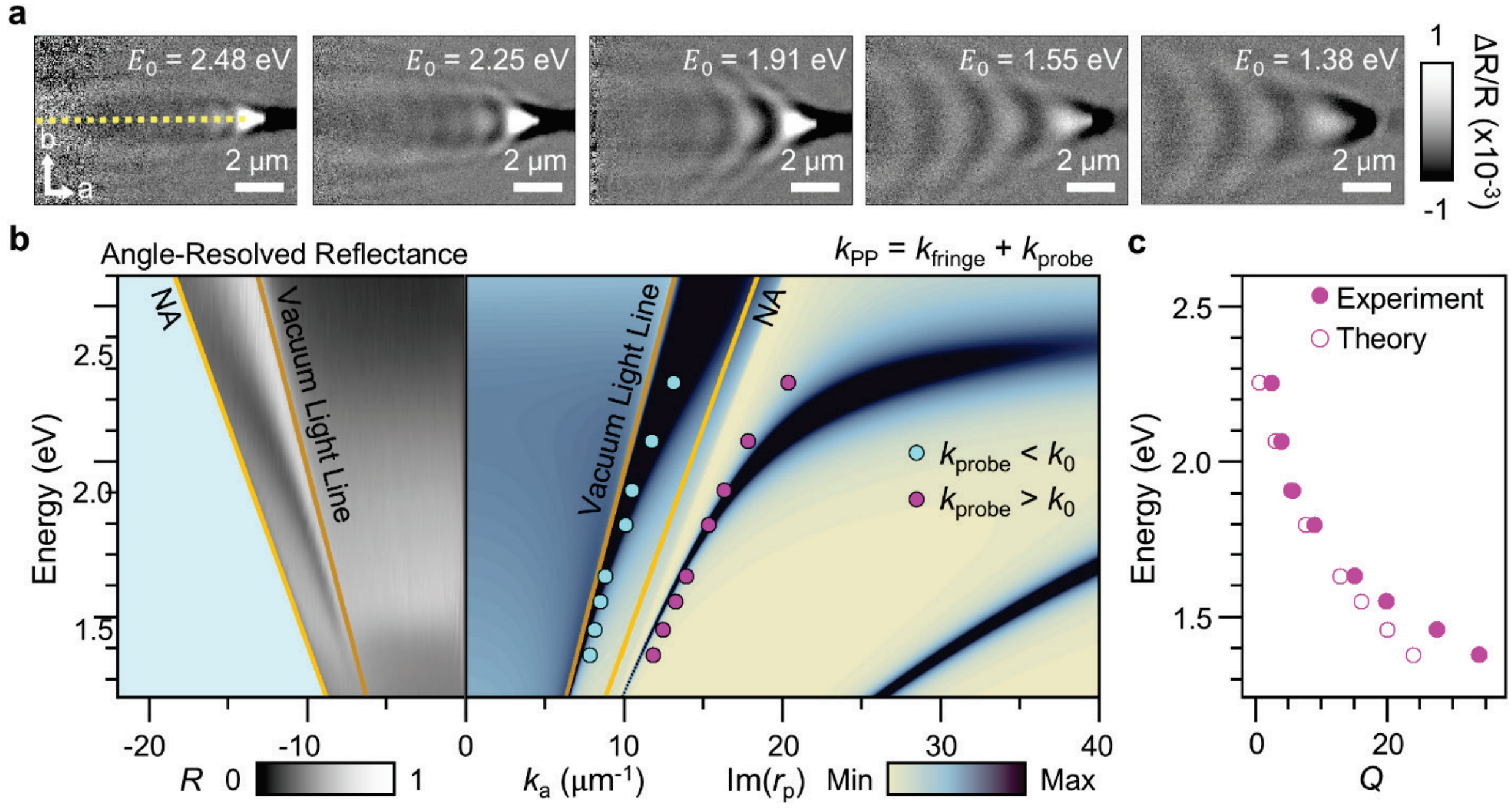

**Fig. 4. Plasmon polariton dispersions. a,** Pump-probe images for a $MoOCl_2$ flake ($d$ = 62 nm) at pump energy 1.72 eV and various probe energies ($k_{probe} = 1.38\ k_0$). The dashed line represents the position of the line profile to extract $k_{fringe}$ (Supplementary Fig. 6a). **b,** Experimental angle-resolved reflectance (left) and simulated (right) dispersion along the $a$-axis for a $MoOCl_2$ flake ($d$ = 62 nm). The cyan and magenta circles denote the corrected wavevector, $k_{PP} = k_{fringe} + k_{probe}$, obtained using $k_{probe} < k_0$ and $k_{probe} > k_0$, respectively. The vacuum light line and NA cutoff are labeled. Note that the angle-resolved reflectance (left panel) drops at energies below 1.5 eV due to the limited range of polarizing optics used in the measurement. **c,** $Q$ factor for the same flake from **b** obtained experimentally (filled circles, $k_{probe} > k_0$) and from theory (open circles) for the first HPP mode.

## Outlook

We demonstrate the hyperbolic spin Hall effect in the natural anisotropic van der Waals metal $MoOCl_2$. Owing to strong optical spin-orbit interactions associated with large anisotropy and confinement, hyperbolic polaritons in $MoOCl_2$ exhibit strong helicity-momentum locking. We leverage these interactions to realize on-demand polarization switching of ray-like hyperbolic polariton trajectories with near-perfect selectivity. These polaritons can be used to generate enhanced chiral near fields, a potential route to sought-after cavity-induced symmetry breaking in quantum materials[47]. Key to our discovery is the realization that oblique-incidence interferometric pump-probe microscopy enables the polarization-selective interrogation of hyperbolic modes in the far-field. Using momentum-dependent imaging, we map the complex plasmonic topology of $MoOCl_2$ and capture the transition from elliptical to hyperbolic polaritons. Overall, our findings establish a foundation for harnessing optical spin-orbit interactions at deeply subwavelength scales, unlocking nanophotonic control and advanced imaging schemes powered by natural hyperbolic materials.

## Methods

### Sample preparation

$MoOCl_2$ bulk crystals used for the data shown in the text were synthesized in Berkeley from the same synthetic batch reported in a previous study[22]. Control crystals synthesized at Columbia showed the same behavior, confirming reproducibility across different synthetic batches.
22 mm x 22 mm $SiO_2$ substrates were rinsed with deionized (DI) water, acetone, isopropyl alcohol, and DI water, sequentially, in a sonicator for 15 minutes per solvent. Once dry, the substrates were $O_2$ plasma cleaned for 10 minutes. Bulk $MoOCl_2$ crystals were mechanically exfoliated with Scotch Magic Tape, followed by exfoliation onto thermal release tape. The exfoliation on thermal release tape was adhered to a clean $SiO_2$ substrate and heated at 100°C for 2-3 minutes. We determined flake thickness using atomic force microscopy (Supplementary Fig. 7).

### Oblique-incident pump-probe microscopy

The oblique-incident pump-probe microscopy we employed is based on our prior developments[38,48,49]. For our pump-probe images, we used a 732 nm confocal pump laser (PicoQuant laser diode LDG-D-C-730) and widefield probe laser with selectable wavelength across the visible and near-infrared regimes (SuperK Extreme supercontinuum laser, ~10 nm bandwidth). Both pump and probe were directed to a high numerical-aperture oil-immersion objective (Leica HC Plan Apo 63x, 1.4 NA oil immersion). A polarizer and half-wave plate in the probe excitation path were used to select the incident polarization, and a half-wave plate and a polarizing beam splitter in the imaging path were used select the detection polarization. A pair of quarter-wave plates are placed in the probe excitation and detection paths, respectively, for measurements using circularly polarized probe. The incident angle of the widefield probe is tuned using a mirror positioned one focal length before the probe widefield lens. $k_{\mathrm{probe}}$ is quantified by imaging the probe position in the back focal plane (Supplementary Section 3). All measurements were carried out at room temperature.

### Angle-resolved reflectance

Angle-resolved reflectance spectra were collected using a stabilized tungsten-halogen lamp (Thorlabs SLS201L). The collimated beam was focused onto the sample using the same microscope objective as in the previous section. The angle-resolved spectra were obtained by projecting the back focal plane was onto a vertical slit of a home-built prism spectrometer. In the spectrometer path, we positioned a half-wave plate and polarizing beam splitter, the latter transmitting horizontally polarized light; we used these optics to collect reflectance spectra under p and s polarizations. All steady-state spectra were normalized to the reflectance of a thick silver film. All measurements were carried out at room temperature.

### Transfer matrix method calculation

We calculated the imaginary part of the reflection coefficient, $\mathrm{Im}(r_{\mathrm{p}})$, of the vacuum/$MoOCl_2$/$SiO_2$ multilayer using the 4x4 transfer matrix formalism for anisotropic media developed by Passler and Paarmann[50]. The dielectric constants of $MoOCl_2$ were obtained from recent literature ($\mathrm{Im}(r_{\mathrm{p}})$[21,34] and the calculated quality factor[37]), assuming a diagonal permittivity tensor. The refractive index of $SiO_2$ was assumed to be a constant of 1.52.

**Numerical simulation**

The scattered electric field profiles shown in Fig. 2 were computed using COMSOL Multiphysics 6.1. The simulation geometry was a $SiO_2/MoOCl_2$/vacuum multilayer. To model the optical pump-induced scatterers, a cylindrical region (radius 10 nm) at the center of $MoOCl_2$ film in which the plasma frequency along the *a*-axis was reduced by a factor of 0.6. To isolate the field scattered by the cylindrical scatterer from the background response of the multilayer (e.g., widefield transmission and reflections at the $SiO_2/MoOCl_2$ and $MoOCl_2$/vacuum interfaces), we employed a two-step procedure. First, the model computes the total field excited by an obliquely incident plane wave launched from a periodic port at 67° for a homogeneous $MoOCl_2$ film. Periodic boundary conditions were applied to the lateral walls to model a laterally infinite structure. The computed field, which consists of the transmitted and reflected fields within the multilayer, were then used as the background excitation for a second simulation in a finite domain surrounded by perfectly matched layers (PMLs) to extract the scattered response. As the PMLs do not fully suppress residual reflections from the lateral boundaries of a highly anisotropic material at large incident angles, the finite-domain calculation was performed both with and without the cylindrical scatterer, and we took the difference between them. The differential field is the electric field scattered solely by the pump-induced scatterer.

## Data availability

The data generated in this study are displayed in the manuscript and supplementary information. Raw data files are available from the corresponding authors upon request.

## Code availability

The details of the simulations are described in Supplementary Information. Additional information related to this paper is available from the corresponding authors upon request.

## References

1. Basov, D. N., Fogler, M. M. & García De Abajo, F. J. Polaritons in van der Waals materials. *Science* **354**, (2016).
2. Wang, H. *et al.* Planar hyperbolic polaritons in 2D van der Waals materials. *Nat. Commun.* **15**, 69 (2024).
3. de Abajo, F. J. G. *et al.* Roadmap for photonics with 2D materials. *ACS Photonics* **12**, 3961–4095 (2025).
4. Bliokh, K. Y., Rodríguez-Fortuño, F. J., Nori, F. & Zayats, A. V. Spin–orbit interactions of light. *Nat. Photonics* **9**, 796–808 (2015).
5. Van Mechelen, T. & Jacob, Z. Universal spin-momentum locking of evanescent waves. *Optica* **3**, 118–126 (2016).
6. Yin, X., Ye, Z., Rho, J., Wang, Y. & Zhang, X. Photonic spin hall effect at metasurfaces. *Science* **339**, 1405–1407 (2013).
7. Rodríguez-Fortuño, F. J. *et al.* Near-field interference for the unidirectional excitation of electromagnetic guided modes. *Science* **340**, 328–330 (2013).
8. Lin, J. *et al.* Polarization-controlled tunable directional coupling of surface plasmon polaritons. *Science* **340**, 331–334 (2013).
9. Kapitanova, P. V *et al.* Photonic spin Hall effect in hyperbolic metamaterials for polarization-controlled routing of subwavelength modes. *Nat. Commun.* **5**, 3226 (2014).

10. Ling, X. *et al.* Recent advances in the spin Hall effect of light. *Reports Prog. Phys.* **80**, 066401 (2017).
11. Rechcinska, K. *et al.* Engineering spin-orbit synthetic Hamiltonians in liquid-crystal optical cavities. *Science* **366**, 727–730 (2019).
12. Kim, M. *et al.* Spin Hall Effect of Light: From Fundamentals To Recent Advancements. *Laser Photonics Rev.* **17**, 2200046 (2023).
13. Liang, J. *et al.* Polariton spin Hall effect in a Rashba–Dresselhaus regime at room temperature. *Nat. Photonics 2024 184* **18**, 357–362 (2024).
14. Xiang, B. *et al.* Optical spin hall effect in exciton-polariton condensates in lead halide perovskite microcavities. *J. Chem. Phys.* **160**, 161104 (2024).
15. Nemilentsau, A., Stauber, T., Gómez-Santos, G., Luskin, M. & Low, T. Switchable and unidirectional plasmonic beacons in hyperbolic two-dimensional materials. *Phys. Rev. B* **99**, 201405 (2019).
16. Hu, G., Krasnok, A., Mazor, Y., Qiu, C.-W. & Alù, A. Moiré hyperbolic metasurfaces. *Nano Lett.* **20**, 3217–3224 (2020).
17. Pian, C. *et al.* Hyperbolic phonon polaritons-induced photonic spin Hall effect in an α-MoO3 thin film. *Appl. Phys. Lett.* **124**, 121602 (2024).
18. Dai, S. *et al.* Tunable phonon polaritons in atomically thin van der Waals crystals of boron nitride. *Science* **343**, 1125–1129 (2014).
19. Li, P. *et al.* Infrared hyperbolic metasurface based on nanostructured van der Waals materials. *Science* **359**, 892–896 (2018).
20. Ma, W. *et al.* In-plane anisotropic and ultra-low-loss polaritons in a natural van der Waals crystal. *Nature* **562**, 557–562 (2018).
21. Venturi, G., Mancini, A., Melchioni, N., Chiodini, S. & Ambrosio, A. Visible-frequency hyperbolic plasmon polaritons in a natural van der Waals crystal. *Nat. Commun.* **15**, 9727 (2024).
22. Ruta, F. L. *et al.* Good plasmons in a bad metal. *Science* **387**, 786–791 (2025).
23. Hillenbrand, R., Abate, Y., Liu, M., Chen, X. & Basov, D. N. Visible-to-THz near-field nanoscopy. *Nat. Rev. Mater.* **10**, 285–310 (2025).
24. Kurman, Y. *et al.* Spatiotemporal imaging of 2D polariton wave packet dynamics using free electrons. *Science* **372**, 1181–1186 (2021).
25. Ghosh, A., Raab, C., Spellberg, J. L., Mohan, A. & King, S. B. Direct visualization of visible-light hyperbolic plasmon polaritons in real space and time. at https://doi.org/10.48550/arXiv.2506.13719 (2025).
26. Li, Y. *et al.* Broadband near-infrared hyperbolic polaritons in $MoOCl_2$. *Nat. Commun.* **16**, 6172 (2025).
27. Zhang, Y. *et al.* Manipulating hyperbolic plasmon polaritons at near-infrared in an anisotropic van der Waals crystal. *Nano Lett.* **25**, 15534–15541 (2025).
28. Abbe, E. Beiträge zur Theorie des Mikroskops und der mikroskopischen Wahrnehmung. *Arch. Mikrosk. Anat.* **9**, 413–468 (1873).
29. Huang, B., Bates, M. & Zhuang, X. Super-resolution fluorescence microscopy. *Annu. Rev. Biochem.* **78**, 993–1016 (2009).
30. Jost, A. & Heintzmann, R. Superresolution multidimensional imaging with structured illumination microscopy. *Annu. Rev. Mater. Res.* **43**, 261–282 (2013).
31. Hao, X., Kuang, C., Li, Y. & Liu, X. Evanescent-wave-induced frequency shift for optical superresolution imaging. *Opt. Lett.* **38**, 2455–2458 (2013).

32. Wicker, K. & Heintzmann, R. Resolving a misconception about structured illumination. *Nat. Photonics* **8**, 342–344 (2014).
33. Wang, Z. *et al.* Fermi liquid behavior and colossal magnetoresistance in layered $MoOCl_2$. *Phys. Rev. Mater.* **4**, 41001 (2020).
34. Zhao, J. *et al.* Highly anisotropic two-dimensional metal in monolayer $MoOCl_2$. *Phys. Rev. B* **102**, 245419 (2020).
35. Zhang, Y., Lin, L.-F., Moreo, A. & Dagotto, E. Orbital-selective Peierls phase in the metallic dimerized chain $MoOCl_2$. *Phys. Rev. B* **104**, L060102–L060102 (2021).
36. Gao, H., Ding, C., Sun, L., Ma, X. & Zhao, M. Robust broadband directional plasmons in a $MoOCl_2$ monolayer. *Phys. Rev. B* **104**, 205424 (2021).
37. Melchioni, N. *et al.* Giant optical anisotropy in a natural van der Waals hyperbolic crystal for visible light low-loss polarization control. *ACS Nano* **19**, 25413–25421 (2025).
38. Xu, D. *et al.* Ultrafast imaging of polariton propagation and interactions. *Nat. Commun.* **14**, 3881 (2023).
39. Lindfors, K., Kalkbrenner, T., Stoller, P. & Sandoghdar, V. Detection and Spectroscopy of Gold Nanoparticles Using Supercontinuum White Light Confocal Microscopy. *Phys. Rev. Lett.* **93**, 037401 (2004).
40. Ginsberg, N. S., Hsieh, C. L., Kukura, P., Piliarik, M. & Sandoghdar, V. Interferometric scattering microscopy. *Nat. Rev. Methods Prim. 2025 51* **5**, 23- (2025).
41. Li, P. *et al.* Hyperbolic phonon-polaritons in boron nitride for near-field optical imaging and focusing. *Nat. Commun.* **6**, 7507 (2015).
42. Dai, S. *et al.* Subdiffractional focusing and guiding of polaritonic rays in a natural hyperbolic material. *Nat. Commun.* **6**, 6963 (2015).
43. Wu, G., Wan, J. H., Qian, C. & Liu, X. W. Plasmonic scattering interferometric microscopy: decoding the dynamic interfacial chemistry of single nanoparticles. *Acc. Chem. Res.* **58**, 3341–3352 (2025).
44. Hitzelhammer, F., Dong, J., Hohenester, U. & Juffmann, T. When higher resolution reduces precision: quantum limits of off-axis interferometric scattering microscopy. at https://doi.org/10.48550/arXiv:2510.03034 (2025).
45. Ni, X. *et al.* Observation of directional leaky polaritons at anisotropic crystal interfaces. *Nat. Commun.* **14**, 2845 (2023).
46. Huttunen, M. J., Abbas, A., Upham, J. & Boyd, R. W. Label-free super-resolution with coherent nonlinear structured-illumination microscopy. *J. Opt.* **19**, 085504 (2017).
47. Hübener, H. *et al.* Engineering quantum materials with chiral optical cavities. *Nat. Mater. 2020 204* **20**, 438–442 (2020).
48. Delor, M., Weaver, H. L., Yu, Q. & Ginsberg, N. S. Imaging material functionality through three-dimensional nanoscale tracking of energy flow. *Nat. Mater.* **19**, 56–62 (2020).
49. Hong, Y., Xu, D. & Delor, M. Exciton delocalization suppresses polariton scattering. *Chem* **0**, 102759 (2025).
50. Paarmann, A. & Passler, N. C. Generalized $4 \times 4$ matrix formalism for light propagation in anisotropic stratified media: study of surface phonon polaritons in polar dielectric heterostructures. *J. Opt. Soc. Am. B* **34**, 2128–2139 (2017).

## Acknowledgements

Work on the plasmonic properties of $MoOCl_2$ was primarily supported by Programmable Quantum Materials, an Energy Frontier Research Center funded by the US Department of Energy (DOE), Office of Science, Basic Energy Sciences, under award DE-SC0019443. Pump-probe and linear reflectance measurements are supported by the National Science Foundation under award CHE-2203844, and by the Beckman Young Investigator award from the Arnold and Mabel Beckman Foundation (M.D.). The authors thank Stephen Sanders for helpful discussions regarding the simulations and Ryo Mizuta Graphics for providing the 3D optical component illustrations in Blender used in the manuscript figures. We acknowledge the use of the Columbia University shared facilities and instrumentation, specifically the atomic force microscope, supported by the National Science Foundation through the Columbia Nano Initiative and the Materials Research Science and Engineering Center DMR-2011738. The Flatiron Institute is a division of the Simons Foundation.

## Author contributions

A.S.D., F.T., D.N.B., and M.D. conceptualized and designed this study. A.S.D. and F.T. prepared samples, performed the measurements and simulations, and analyzed the data, with technical help from D.X., I.L., and N.B. D.B. and J.G.A synthesized the $MoOCl_2$ crystals used for the data shown in the text. E.L., C.N., and X.R. synthesized control $MoOCl_2$ crystals for comparison. A.S.D., F.T., D.N.B., and M.D. wrote the manuscript, with input from all authors.

## Competing interests

The authors declare no competing interests.

**Supplementary Information for:**

# On-demand steering of hyperbolic chiral polaritons

Andrea S. Dai[1,#], Fuyang Tay[1,2,#], Ding Xu[1], Inki Lee[1], Noah Bussell[1], Daria Balatsky[3,4], Francesco L. Ruta[2,5], Emma Lian[1], Colin Nuckolls[1], Xavier Roy[1], James G. Analytis[4,6,7], Andrew J. Millis[2,8], D. N. Basov[2*], Milan Delor[1*]

[1]Department of Chemistry, Columbia University, New York, NY, USA.
[2]Department of Physics, Columbia University, New York, NY, USA.
[3]Department of Chemistry, University of California, Berkeley, CA, USA.
[4]Department of Physics, University of California, Berkeley, CA, USA.
[5]Department of Applied Physics and Applied Mathematics, Columbia University, New York, NY, USA.
[6]CIFAR Quantum Materials, Toronto, ON, Canada.
[7]Kavli Energy Nanoscience Institute, Berkeley, CA, USA.
[8]Center for Computational Quantum Physics, Flatiron Institute, New York, NY, USA.

[#]These authors contributed equally to this work.

[*]Corresponding authors. db3056@columbia.edu; milan.delor@columbia.edu

**Table of Contents**

## Supplementary Section 1. Point spread functions of oblique illumination microscopy

The PSF (point spread function) of an obliquely incident probe can be described as an Airy disk shifted away from the origin in Fourier space. The Fourier-space shift can be implemented by multiplying the Airy function by a phase factor, $e^{ikx}$. Supplementary Fig. 1 displays the expected PSF in both real and Fourier space for various $k_{\mathrm{probe}}$ (tilted along the $x$ axis) at 1.91 eV with NA = 1.4. As $k_{\mathrm{probe}}$ increases, the PSF evolves from an Airy disk to elliptical and ultimately parabolic features. Supplementary Section 4 shows that this PSF can be readily deconvolved from the hyperbolic signal. The PSF exhibits opposite phases between $x < 0$ and $x > 0$, as seen in the imaginary component of PSF, Im(PSF). The Fourier transform of the PSF corresponds to the microscope's passband in Fourier space.

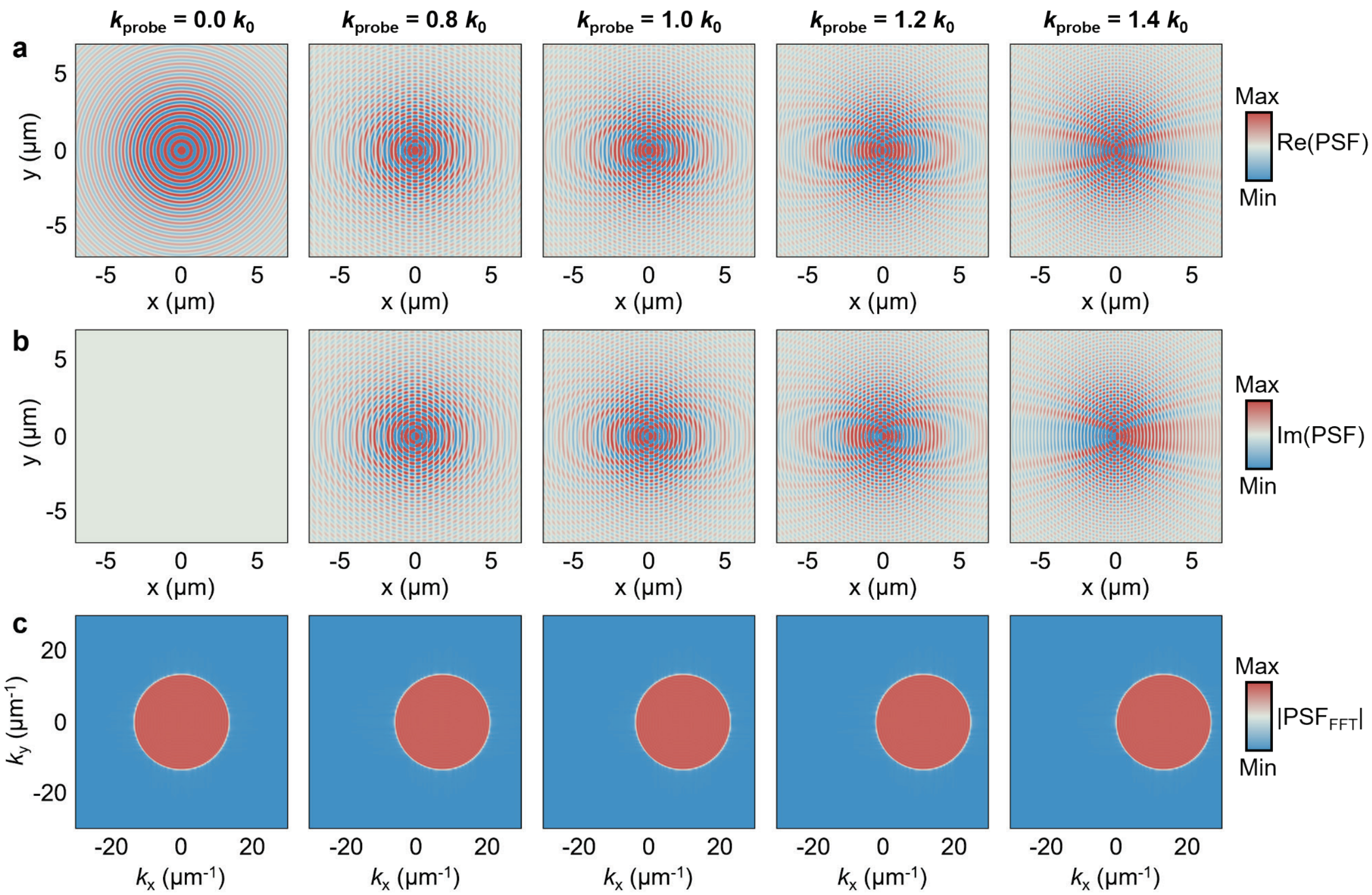

**Supplementary Fig. 1 PSF of oblique illumination microscopy for various $k_{\mathrm{probe}}$ at 1.91 eV with NA = 1.4.** The (**a**) real and (**b**) imaginary components of the PSF as a function of $k_{\mathrm{probe}}$. **c,** Fourier transform of the PSF.

## Supplementary Section 2. Pump-probe images with different oblique incidence orientations

Supplementary Fig. 2 shows pump-probe images using a p-polarized probe obliquely incident on the sample from different orientations with respect to the crystal axes. The hyperbolic 'V'-shaped periodic fringes only appear when the probe is tilted along the *a*-axis (horizontal).

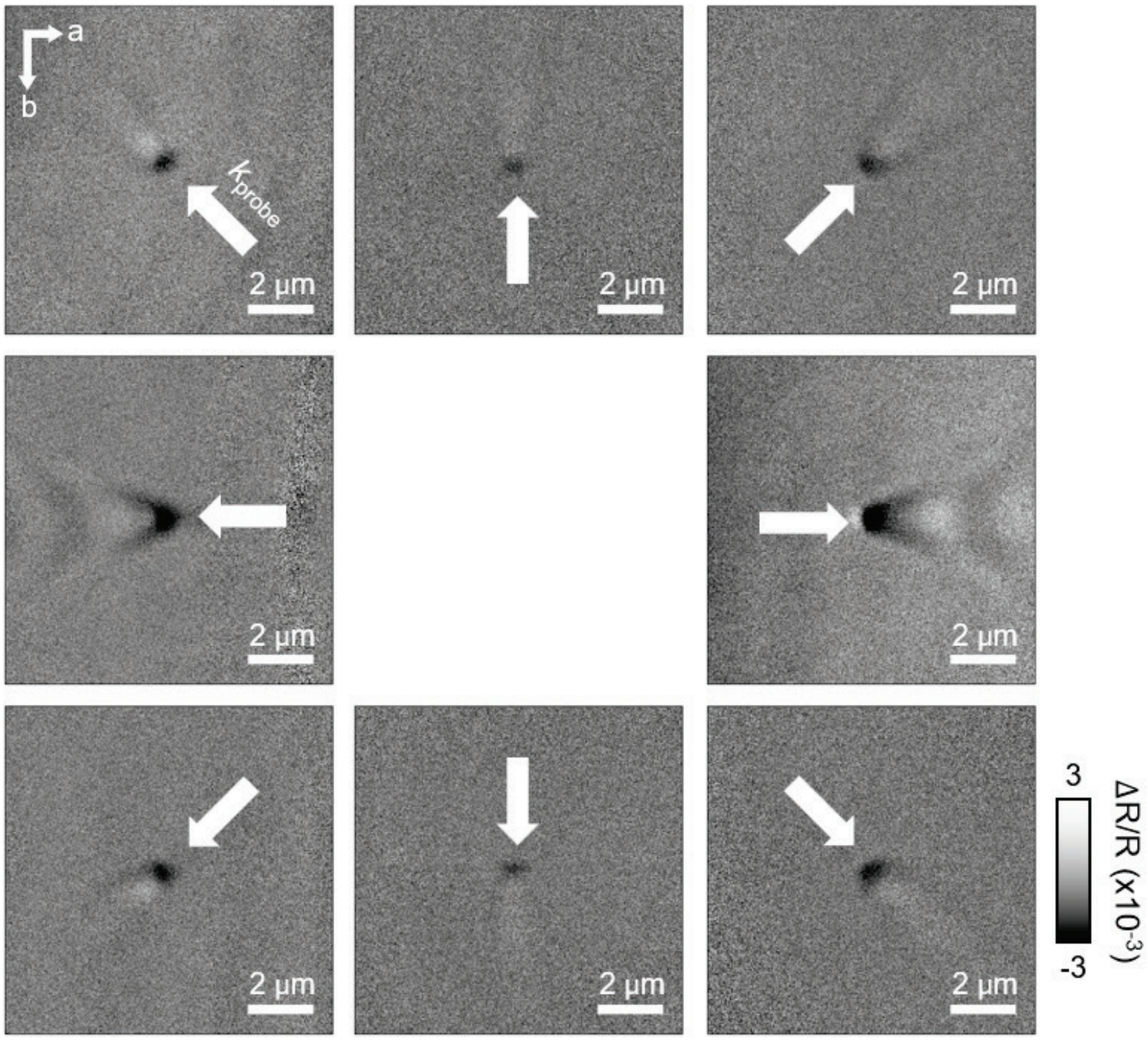

**Supplementary Fig. 2 Pump-probe images acquired with a p-polarized probe with oblique incidence angle oriented along different directions.** The images are recorded at 0.4 ns pump-probe delay on a 62 nm thick $MoOCl_2$ flake. The white arrows indicate the oblique probe incidence angles. The pump and probe energies are 1.72 eV and 1.59 eV, respectively.

## Supplementary Section 3. Extraction of $k_{probe}$

The $k_{probe}$ values were obtained directly from angle-resolved reflectance measurements through back focal (Fourier) plane imaging. The $k$ in the back focal plane was first calibrated using total internal reflection lines and the objective's NA limit. The obliquely incident probe in the back focal plane was then recorded on the camera, and its intensity profile was fit with a Gaussian function to extract the peak position (and thus $k_{probe}$). For instance, Supplementary Fig. 3 shows the Gaussian fits to the probe's profile for $k_{probe}$ in Fig. 3 of the main text.

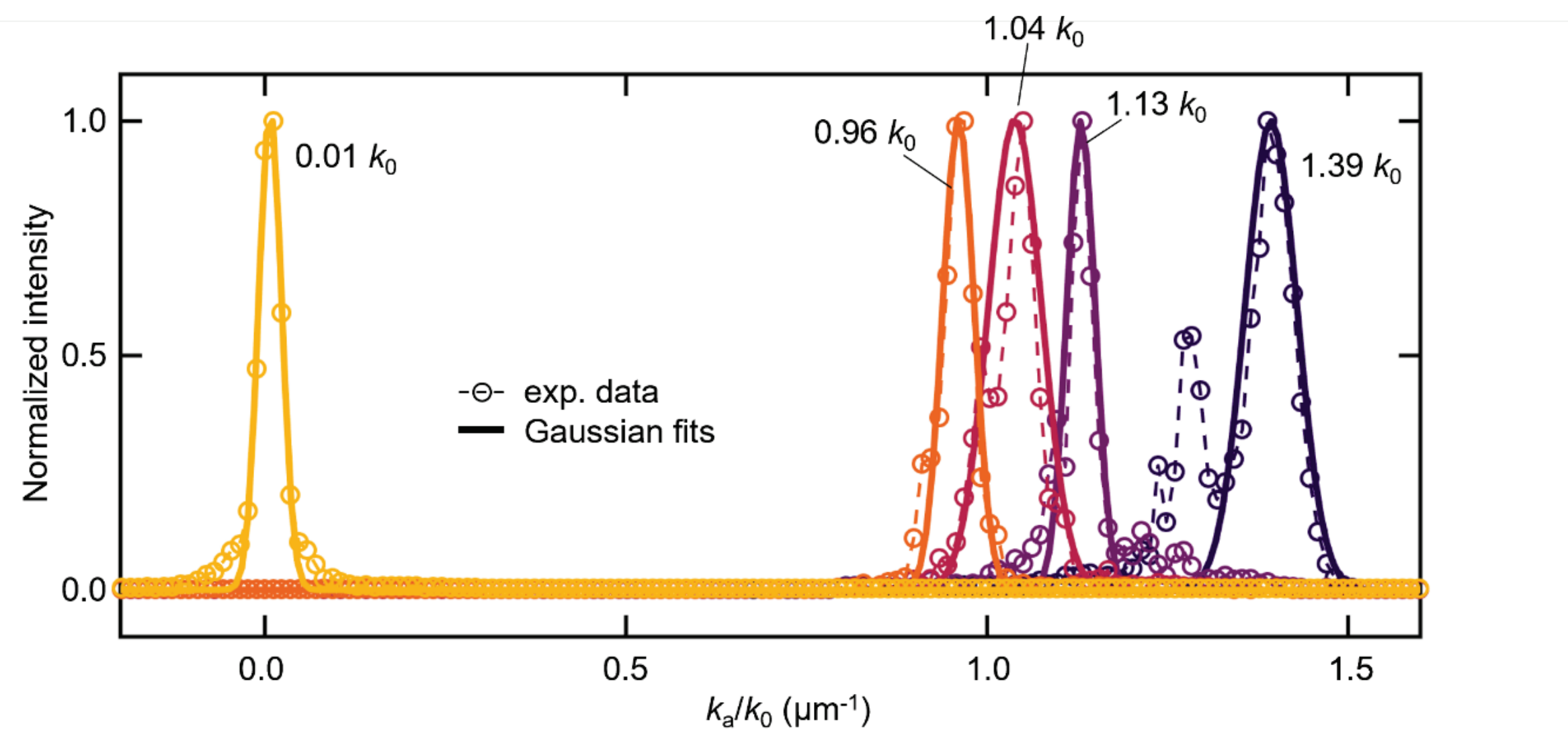

**Supplementary Fig. 3 Extraction of $k_{probe}$ from back focal plane imaging.** Normalized intensity line profiles of the obliquely incident, collimated probe (open circles with dashed lines) used in Fig. 3 of the main text and corresponding Gaussian fits (solid lines). The fitted peak positions yield $k_{probe}$.

## Supplementary Section 4. Deconvolving the PSF from the pump-probe signal

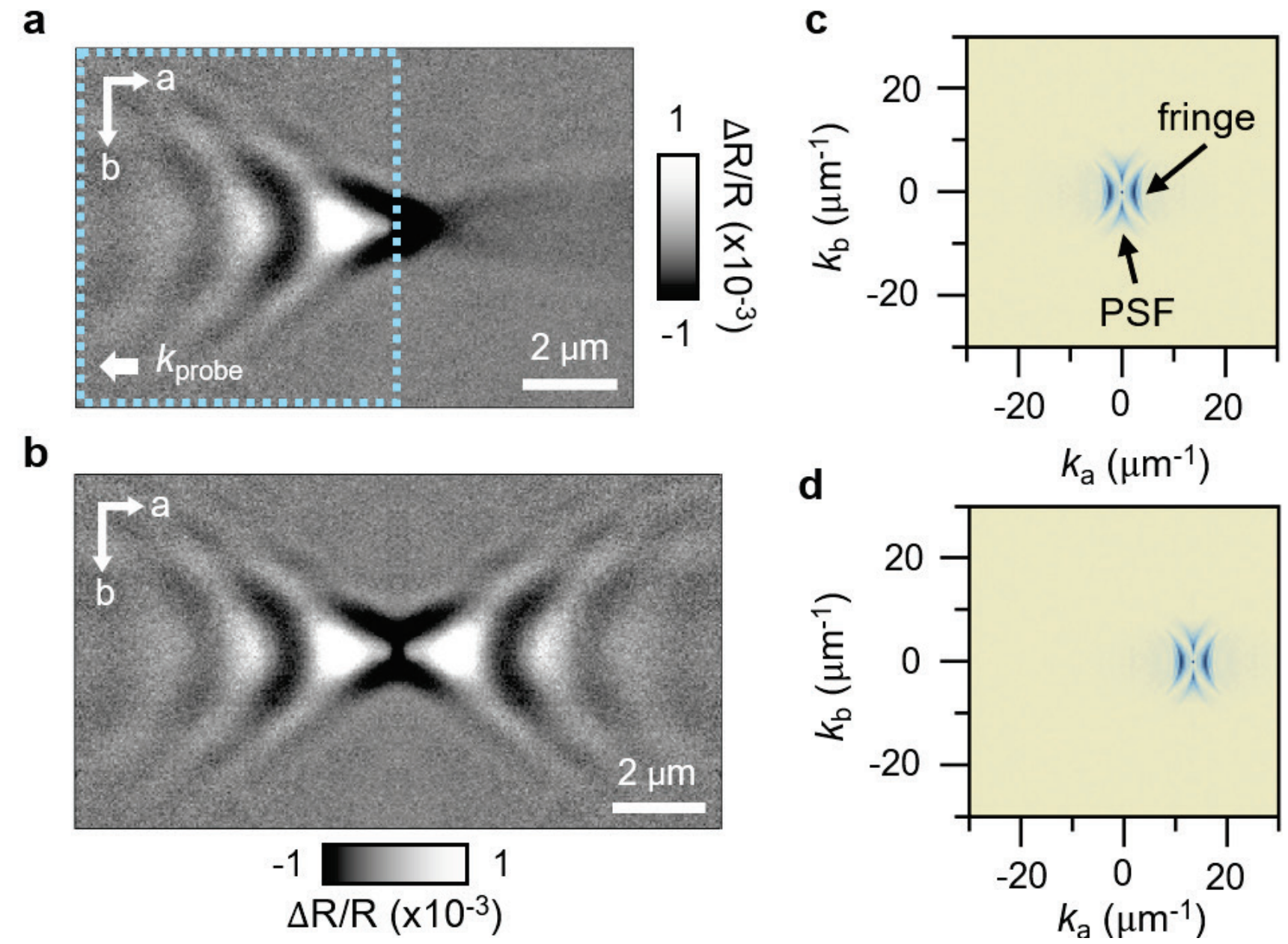

**Supplementary Fig. 4 Fourier transform processing procedures to convert Fig. 3b panels into Fig. 3c. a,** Pump-probe image of a 62-nm-thick flake at pump/probe energies 1.72 eV/1.91 eV and $k_{probe}$ = 1.39$k_0$ (same as Fig. 3b, right panel). The blue rectangle selects the forward direction of the signal to be Fourier transformed by the following procedure: **b,** mirror the selection in **a**; **c,** fast Fourier transform the image in **b**, which results in distinct features in *k*-space for the PSF and the periodic fringe pattern (labeled); **d,** apply a frequency shift along the *a*-axis with the value of $k_{probe}$.

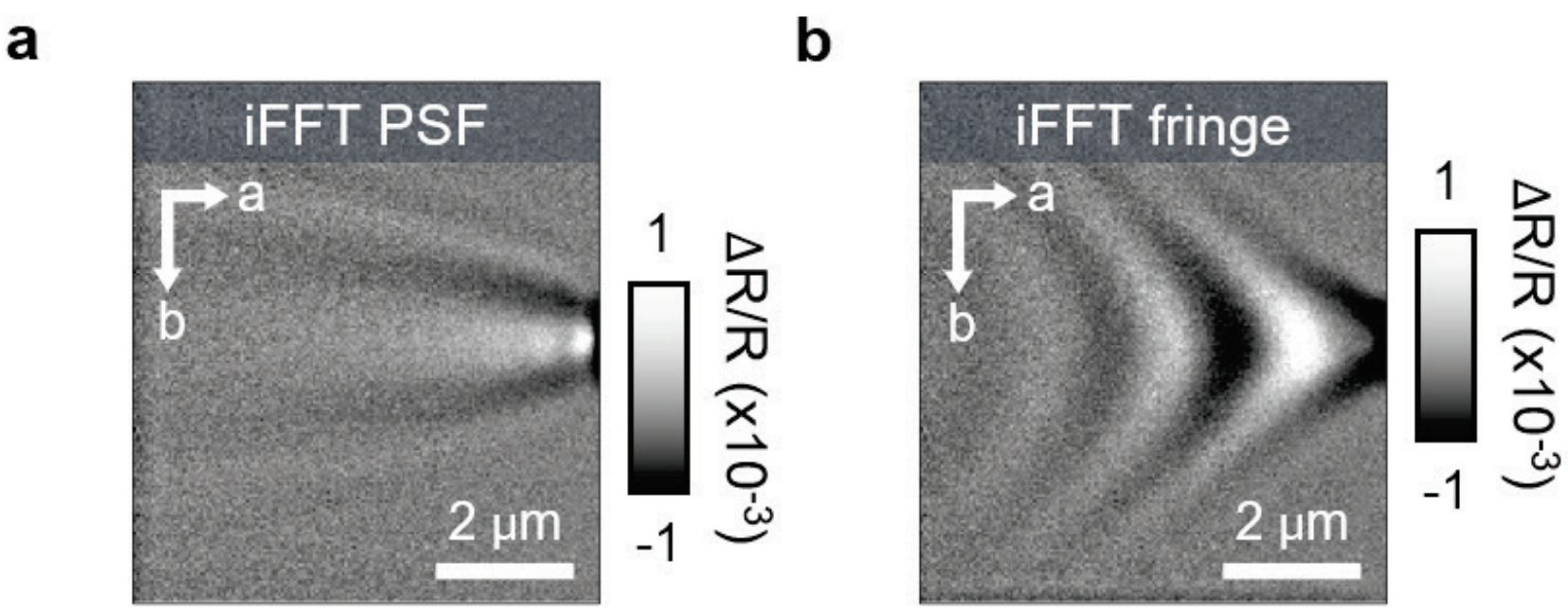

**Supplementary Fig. 5 Fourier filtering of pump-probe images.** Inverse Fourier transform (iFFT) of Supplementary Fig. 4c after filtering out the features attributed to **a** the PSF and **b** the plasmonic fringes in *k*-space, respectively. All images shown in the main text are raw, unprocessed images where the fringes remain convolved with the PSF.

## Supplementary Section 5. Two methods for extracting $k_{fringe}$

The values of $k_{\mathrm{fringe}} + k_{\mathrm{probe}}$ may be extracted via two methods. In the first method, we perform 1D fast Fourier transform (FFT) on a line cut across the $a$-axis (illustrated as a dashed line in Fig. 4a of the main text), which yields distinct peaks for the PSF and plasmon fringes, as shown in Supplementary Section 4. The FFT peak corresponding to the pump-probe fringes is converted to $k_{\mathrm{fringe}}$. In the second method, we fit the forward line cut with Equation 1 (Supplementary Fig. 6a), an exponentially damped sinusoidal function:

$$\Delta R/R = A\, e^{-\mathrm{Im}(k_{\mathrm{fringe}})x} \sin\left(\mathrm{Re}(k_{\mathrm{fringe}})x + \phi\right)/\sqrt{x} + B\,, \tag{1}$$

where $A$, $B$, *and* $\phi$ are constant fitting parameters. The extracted values from both methods are plotted in Supplementary Fig. 6b and show excellent agreement.

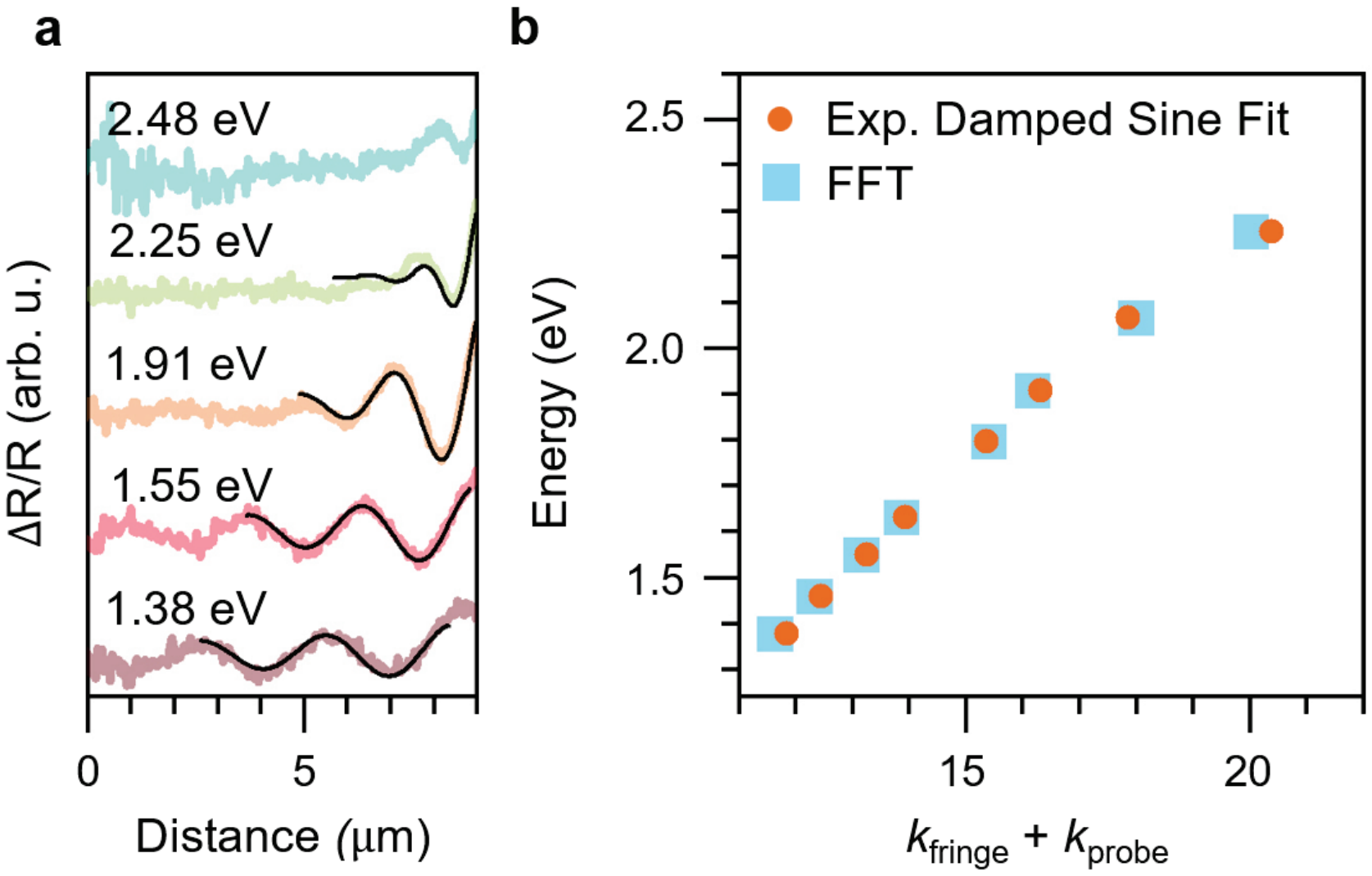

**Supplementary Fig. 6 Comparison of methods for extracting $k_{\mathrm{fringe}} + k_{\mathrm{probe}}$ when $k_{\mathrm{probe}} > k_0$.** **a,** Linecut profiles along the dashed line in Fig. 4a of the main text, labeled with the corresponding probe energy. Black solid lines are exponentially damped sinusoidal fits to the data. **b,** Summary of $k_{\mathrm{fringe}} + k_{\mathrm{probe}}$ determined via exponentially damped sinusoidal fitting (orange circles) and fast Fourier transform (blue squares).

## Supplementary Section 6. Thickness dependence

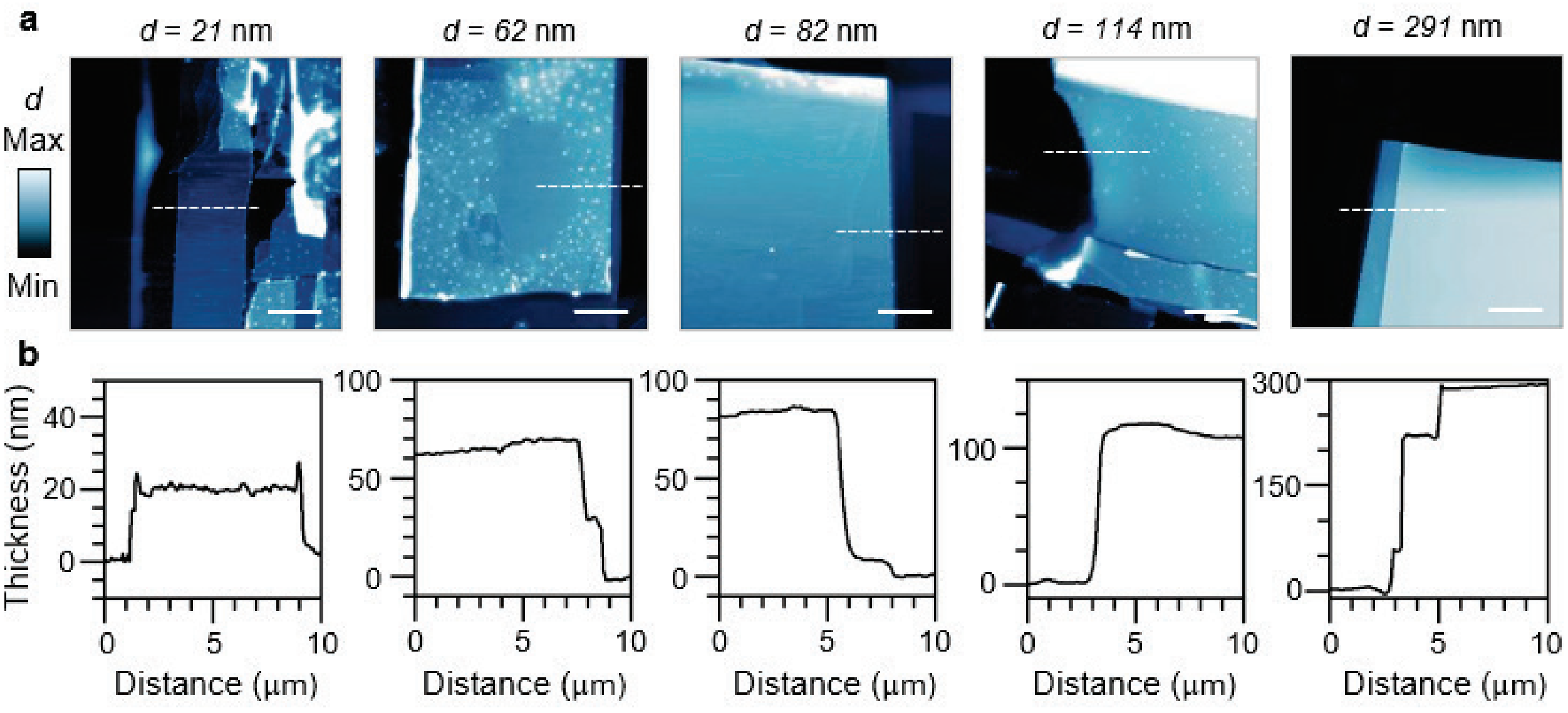

**Supplementary Fig. 7 $MoOCl_2$ flakes with different thicknesses. a**, AFM images of $MoOCl_2$ flakes of various thicknesses, *d*. The color scaling is different for each thickness, adjusted for image clarity. **b,** Line profiles corresponding to the dashed white lines in **a**. Scale bars, 5μm.

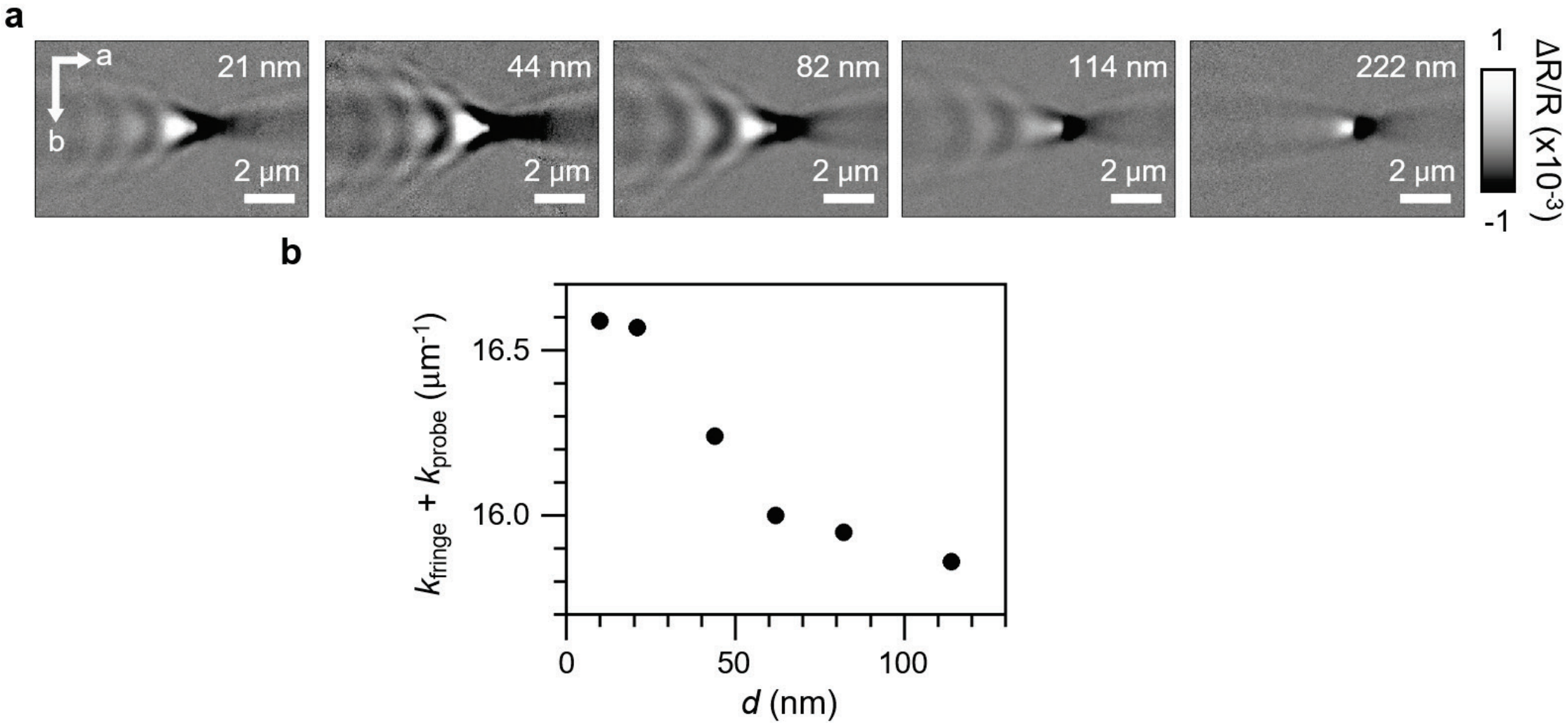

**Supplementary Fig. 8 Thickness dependence of extracted $k_{\text{fringe}} + k_{\text{probe}}$. a**, Pump-probe images for flakes of various *d*, measured at pump/probe energies of 1.72 eV/1.91 eV. Scale bars, 2 μm. **b**, $k_{\text{fringe}} + k_{\text{probe}}$ extracted from measurements with $k_{\text{probe}} = 1.39k_0$ versus *d*.